\documentclass{elsart}

\usepackage{natbib}

 \usepackage{graphicx}

\usepackage{amssymb}

\begin{document}

\begin{frontmatter}



\title{Black-Hole Transients : From QPOs to Relativistic Jets}


\author{T. Belloni}

\address{INAF - Osservatorio Astronomico di Brera, Via E. Bianchi 46,
	I-23807 Merate, Italy}

\begin{abstract}
Due to the impressive amount of new data provided by the RXTE satellite 
in the past decade, our knowledge of the phenomenology of accretion onto 
black holes has increased considerably. In particular, it has been possible 
to schematize the outburst evolution of transient systems on the basis of 
their spectral and timing properties, and link them to the ejection of 
relativistic jets as observed in the radio. Here, I present this scheme, 
concentrating on the properties of the quasi-periodic oscillations observed 
in the light curves and on the link with jet ejection.
\end{abstract}


\end{frontmatter}

\section{Introduction}
\label{}

The RossiXTE mission has changed our view of the high-energy
emission of Black Hole Transients (BHT),
leading to a new, complex picture which is difficult to
interpret. At the same time, a clear connection between X-ray and
radio properties has been found (see Fender 2005, Fender et al.
2004). In the following, I
present briefly the state paradigm that is now emerging, based
on a large wealth of RossiXTE data from bright transient sources
and its connection with jet ejection
(see Belloni et al. 2005; Homan \& Belloni 2005; Fender et al. 2004). 
Although a full discussion would necessitate a much longer paper, 
I  will simplify the picture as much as possible, concentrating on the
most important issues. In particular, I will stress the connection between
fast-timing properties and presence of a jet.

\section{Black Hole States}

The results of detailed timing and color/spectral analysis of the RossiXTE data
of bright BHTs have evidenced a very wide range of phenomena which are
difficult to categorize. Nevertheless, it is useful to identify distinct
states. Based on the variability and spectral behavior and the transitions
observed in different energy bands, I consider the following states in addition
to a quiescent state (see Homan \& Belloni 2005; Belloni et al. 2005; see also
Casella et al. 2004,2005a for the description of the different QPO types):

\begin{itemize}

\item Low/Hard State (LS): this state is the one associated to relatively low
values of the accretion rate, i.e. lower than in the other bright states, but
it can be observed at all luminosities. The
energy spectrum is hard and the fast time variability is dominated by a
strong ($\sim$30\% fractional rms) band-limited noise. Sometimes, low
frequency QPOs are observed. The characteristic frequencies detected in the
power spectra follow broad-range correlations (see Belloni
et al. 2002). In this state, flat-spectrum radio emission is observed,
associated to compact jet ejection (see Gallo et al. 2003;
Fender et al. 2004).

\item Hard Intermediate State (HIMS): the energy spectrum is
softer than in the LS, with evidence for a soft thermal disk component. The
power spectra feature band-limited noise with characteristic frequency
higher than the LS and usually a rather strong 0.1-15 Hz type-C QPO
(see e.g. Casella et al. 2005a). The
frequencies of the main components detected in the power spectra
extend the broad correlations mentioned for the LS.
The radio emission shows a slightly steeper spectrum (Fender et al. 2004). 
Just before the transition to the SIMS (see below),
Fender et al. (2004) suggested that the
jet velocity increases rapidly, giving origin to a fast relativistic jet.

\item Soft Intermediate State (SIMS): here the energy spectrum is systematically
softer than the HIMS. The disk component dominates the flux. No strong
band-limited noise is observed, but transient type-A and type-B QPOs, the
frequency of which spans only a limited range. No core radio emission is
detected. The few instances of high-frequency QPOs in BHT were observed in this
state (see e.g. Morgan et al. 1997; Homan et al. 2001,2003; Cui et al. 2000;
Remillard et al. 1999).

\item High/Soft State (HS): the energy spectrum is very soft and
strongly dominated by a thermal disk component. Only weak power-law noise is observed
in the power spectrum. No core radio emission is detected
(see Fender et al. 1999; Fender 2005)

\end{itemize}

\begin{figure}
  \includegraphics[width=15.4 true cm]{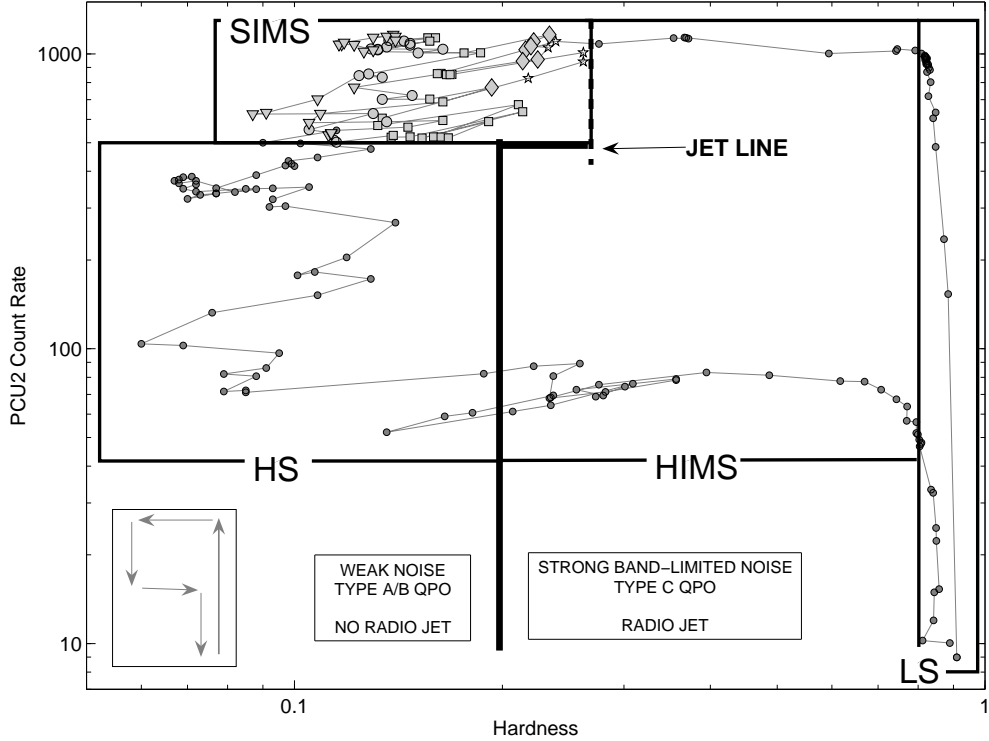}
  \caption{Hardness-Intensity diagram of the 2002/2003 outburst of
        GX 339-4 as observed by the RXTE PCA. The lines mark
        the four source states described in the text. The dashed line
	shows the position of the `jet line'. The thick line, intended
	as a prolongation of the jet line, indicates
	the transition line that marks the presence/absence of strong
	band-limited noise. The inset on the
        lower left shows the general time evolution of the outburst
        along the 'q'-shaped pattern. From Belloni et al. (2005).}
\end{figure}

This classification originates from the analysis of the time evolution
in BHTs. The states described above are defined also in
terms of their transitions, which need to be taken into account. 
A sketch
of the evolution of the 2002/2003 outburst of GX 339-4
is ideal to show the transitions (see Homan \& Belloni 2005;
Belloni et al. 2005).
Figure 1 shows the outburst in a Hardness-Intensity Diagram (HID): the x-axis
shows the X-ray hardness and the y-axis the detected count rate (from
Belloni et al. 2005). The general direction of the time evolution of the
outburst is shown by the arrows in the bottom left corner. As described
in detail below, the position of the boundaries between
different states are absolutely not arbitrary, but are based on the presence of
sharp changes in the observed properties. 

\begin{figure}
  \includegraphics[width=15.4 true cm]{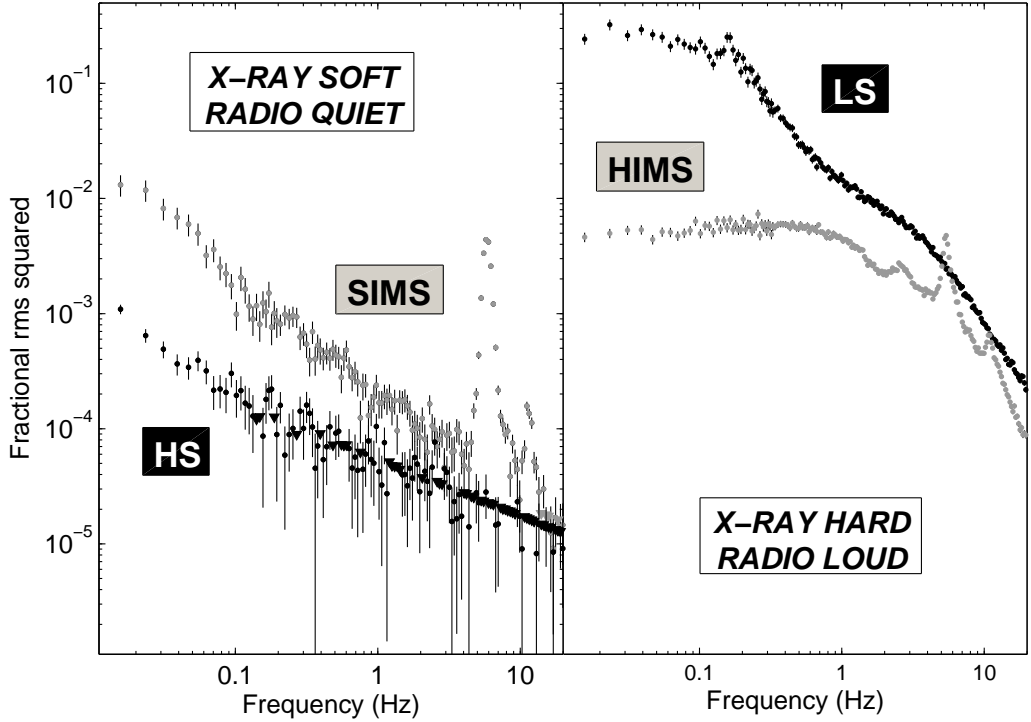}
  \caption{Sample PDS from the 2002/2003 outburst of GX 339-4.
  Left panel: PDS from a SIMS observation (gray points) and average PDS from the
  HS (black points). Right panel: PDS from a LS (gray points) and a HIMS
  observation (black points). The difference in the level of variability between
  the two panels is evident. Data from Belloni et al. (2005)}
\end{figure}

\section{State transitions and jet ejection}

The states outlined above are defined in terms of their transitions, unlike
other proposed state-classifications (see McClintock \& Remillard 2005).
However, as I will show, some transitions appear to be more relevant than
others.

\begin{itemize}

\item {\it LS to HIMS:} the position of this transition in the HID can be
debated, as the color and timing properties change rather smoothly. However,
Homan et al (2005) showed that abrupt changes in the infrared/X-ray correlation
are observed in correspondence of a particular date, which provides a strong
mark for the position of the transition. The IR/X-ray correlation is positive
before the transition, then switches to negative after the transition.
The timing properties do show evidence of changes (see Belloni et al.
2005), but in a rather smooth way.

\item {\it HIMS to SIMS:} here the transition is very sharp. The power spectra
change from band-limited noise plus type-C QPO to red noise plus type-A/B QPO on
a time scale that can be as short as a few seconds (see Nespoli et al. 2003,
Casella et al. 2004). Color changes are not that marked, and indeed from 
INTEGRAL data of the 2004/2005 outburst of GX 339-4 there is
evidence that during such a transition the hard spectral component changes
smoothly (Belloni et al. 2005, in prep.).

\item {\it SIMS to HS:} after the transition line in Fig. 1, all observations do
not show any evidence of features in the power spectrum. Although some SIMS
observations also  appear very quiet, it is remarkable that below a certain
source rate, no more signal is detected. Averaging all observations, a weak
power law is observable in the power spectrum (see Fig. 2).

\item {\it HS to HIMS:} the line in Fig. 1 clearly marks the appearance of
strong band-limited noise (plus type-C QPOs). The three observations that moved
back to the left of this line do not have this noise, indicating that there is
indeed a color boundary. 

\item {\it HIMS to LS:} as in the case of the reverse transition, the color and
timing properties do not show strong discontinuities in their evolution. The
transition line is placed at the same color as the reverse transition.

\end{itemize}

In the model of Fender et al.  (2004), the HIMS--SIMS transition
corresponds to the `jet line', which marks the time of the sudden quenching 
of a rapidly accelerating radio jet (see also Corbel et al. 2004). 
Notice that in some sources this transition
can be repeated: in GRS 1915+105, it takes place on all time scales, and radio
oscillations are observed to correlate with it (see Eikenberry et al. 1998,
Klein-Wolt et al. 2002). Also in XTE J1859+226, rapid transitions are observed:
it is interesting to notice that the appearance of type-B QPOs, tracers of the
SIMS, appear roughly in correspondence of radio flares (see Casella et al.
2005b).

\section{Two states only?}

The jet line does not extend to the low
part of the diagram, but a discontinuity line can be traced vertically (see
Fig. 1). To the right of this line, marked `radio-loud', where a radio 
emitting jet component exists, there is strong band-limited noise in the power
spectrum, often accompanied by type-C QPOs whose frequencies correlate with
spectral parameters (see Vignarca et al. 2003). 
To the left of this line, marked
`radio-quiet', no strong core radio emission is observed (see Fender et al.
1999) and the power spectrum does {\it not} show strong noise, with the
occasional presence of a type-A/B QPOs, whose frequencies vary only over a small
range (see Casella et al. 2005a). 
Typical power spectra (from GX 339-4) can be seen in Fig. 2.

\begin{figure}
  \includegraphics[width=15.4 true cm]{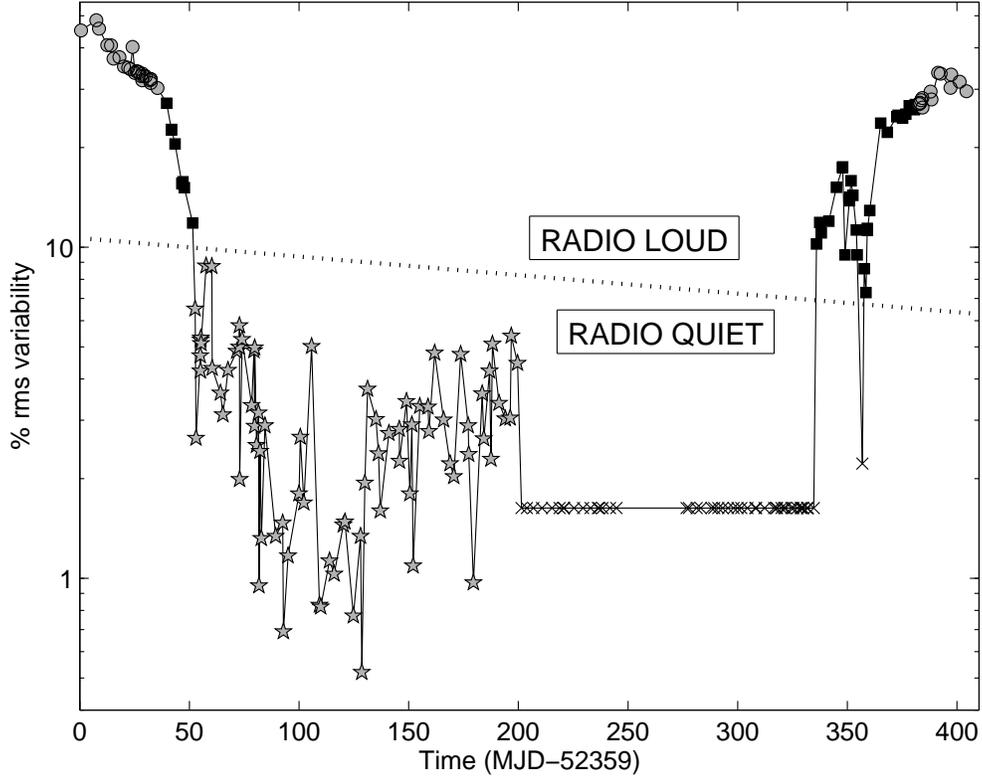}
  \caption{Time evolution of he integrated (0.06-64 Hz)
  fractional rms for the 2002/2003
  outburst of GX 339-4. The different states are marked with different symbols:
  LS (gray circles), HIMS (black circles), SIMS (stars), HS (crosses). Notice
  that the points during the long HS period have all been assigned the average
  rms vale.
  Data from Belloni et al. (2005)}
\end{figure}

The clear separation in timing and radio properties of the hard (LS and HIMS)
and soft (SIMS and HS) states can be seen in Fig. 3, where the time evolution
of the integrated fractional rms of GX 339-4 during.
This indicates that the four states presented above, in terms of physical
conditions, might be reduced to only two: a hard state characterized by
radio-loudness, strong noise components and variable-frequency type-C QPOs, and
a soft one, radio-quiet, with weak power-law noise and transient type-A/B QPOs.


It is important to note that major tracers of these states are low-frequency
QPOs and associated noise components. Type-C QPOs and band-limited noise
components provide characteristic frequencies that trace the evolution of the
accretion during the early and final part of the outburst; in the hard states,
the strong correlations between hard X-ray emission and radio flux led to models
for the jet production and X-ray emission which need to take these timing 
tracers into account (see e.g. Meier 2005). Type-A/B QPOs are much less studied
and their origin is unclear. However, the transient presence of type-B QPOs
could be important for the study of the physical conditions of the accretion
flow as the source crosses the jet line. The small range of long-term
variability of their centroid frequency, its fast short-term variability (see
Nespoli et al. 2003; Belloni et al. 2005) and the absence of band-limited noise
components are key ingredients for their understanding (see Casella et al.
2005a). 

\section{Discussion}

The general picture that can be drawn from the properties discussed above
is the following. In the hard
state, the source is probably jet-dominated, in the sense that the power in
the jet is larger than that in the accretion 
(see Fender et al.
2003). The dominant component in the X-ray range is a thermal hard component,
which is possibly associated to the jet itself (see e.g. Markoff et al. 2003
and Nowak et al. 2005).
The geometrically thin, optically thick, accretion
disk is very soft and has a varying inner radius, so that its contribution to
the X-ray emission changes. This state is characterized by strong
band-limited noise and type-C QPOs, whose characteristic frequencies show
clear correlations between themselves and with spectral parameters (see
Wijnands \& van der Klis 1999;
Belloni et al. 2002; Markwardt et al. 1999; Vignarca et al. 2003).
This state includes the LS and HIMS: the latter is associated to small accretion
disk radii, faster jet ejection, steeper energy spectra and higher
characteristic frequencies. In the HIMS, the corona component (see below)
starts contributing to the X-ray flux (Zdziarski et al. 2001; Rodriguez
et al. 2004).
In the soft states (HS and SIMS), the jet is suppressed 
(radio and X-ray jet components are not observed). The flux is dominated by the
optically thick
accretion disk component, which now has a higher temperature and a small
inner radius, possibly coincident with the innermost stable orbit around
the black hole, but an additional power-law component, with no evidence
of a high-energy cutoff up to $\sim$1 MeV is visible, which I associate
here (generically) to a corona. In the power spectra, no strong band-limited
noise component is observed, and transient QPOs of type A/B are observed,
with frequencies above a few Hz and not much variable on time scales longer than
a few seconds (Casella et al. 2005a).

The above picture involves four observationally separated states, two physical
states and three emission components: the optically thick disk, the thermal hard
component and the corona component. Their contribution to the fast time
variability is markedly different: the disk component does not contribute to it,
the thermal hard component is associated to the band-limited noise and type-C
QPOs, and the corona component is associated to the type-A/B QPOs and the weak
power-law variability observed in the soft states.
A key ingredient is the HIMS--SIMS transition: the transition itself is marked
by the timing properties, but the spectral evolution through it is far from being
clear. Recent INTEGRAL data suggest that 
the transition between the
thermal component being dominant (as in the LS) and the corona component being
visible at high energies (in the HS) is smooth across the jet line
(Belloni et al. 2005, in prep.).

Although timing analysis of the fast variability of BHTs can
give us direct measurements of important parameters of the accretion flow,
up to now we do not have unique models that permit this. Recent results show 
that
a clear association can be made between type-C QPO, strong band-limited noise
and the presence of a relativistic jet. In the framework of unifying models,
these results could play an important role.





\begin{thebibliography}{}


\bibitem{bpk02}
T. Belloni, D. Psaltis, and M. van der Klis, 2002, \emph{ApJ}, 572, 392

\bibitem{b339}
T. Belloni, et al., 2005, \emph{A\&A}, in press (astro-ph/0504577)

\bibitem{cas04}
P. Casella, et al., 2004, \emph{A\&A}, 426, 587

\bibitem{cas05a}
P. Casella, et al., 2005a, \emph{ApJ}, in press (astro-ph/0504318)

\bibitem{cas05b}
P. Casella, et al., 2005b, in \emph{Interacting Binaries: Accretion, Evolution and 
Outcomes}, AIP, in press.

\bibitem{cor04}
S. Corbel, et al., 2004, \emph{ApJ}, 616, 1272

\bibitem{cui00}
P. Cui, et al., 2000, \emph{ApJ}, 535, L123

\bibitem{eik98}
S.S. Eikenberry, et al., \emph{ApJ}, 494, L61

\bibitem{f05}
R.~P. Fender, 2005, in \emph{Compact Stellar X-ray Sources}, Cambridge Univ. Press,
        in press (astro-ph/0303339).
	
\bibitem{fen99}
R.P. Fender, et al., 1999, \emph{ApJ}, 519, L165

\bibitem{fen03}
R.P. Fender, E. Gallo,  and P. Jonker, 2003,  \emph{MNRAS}, 343, L99
	
\bibitem{rpf04}
R.P. Fender, T. Belloni, and E. Gallo, 2004, \emph{MNRAS}, 355, 1105

\bibitem{gfp}
E. Gallo, et al., 2003, \emph{MNRAS}, 344, 60

\bibitem{h01}
J. Homan, et al., 2001, \emph{ApJS}, 132, 377

\bibitem{h03}
J. Homan, et al., 2003, \emph{ApJ}, 586, 1262

\bibitem{h05}
J. Homan, et al., 2005, \emph{ApJ}, 624, 295

\bibitem{hb05}
J. Homan,  and T. Belloni, 2005, in
\emph{From X-ray Binaries to Quasars: Black Hole Accretion on All Mass Scales},
Kluwer, (astro-ph/0412597)

\bibitem{kle02}
M. Klein-Wolt, et al., 2002, \emph{MNRAS}, 331, 745

\bibitem{mar03}
S. Markoff, et al., 2003, \emph{A\&A}, 397, 645

\bibitem{mar99}
C.B. Markwardt, J.H. Swank, and R.E. Taam, 1999, \emph{ApJ}, 513, L37

\bibitem{mcrem}
J.E. McClintock, and R.A. Remillard, 2005, 
in \emph{Compact Stellar X-ray Sources}, Cambridge Univ. Press,
        in press (astro-ph/0306213)
	
\bibitem{mei05}
D.L. Meier, 2005, in
\emph{From X-ray Binaries to Quasars: Black Hole Accretion on All Mass Scales},
Kluwer, (astro-ph/0504511)

\bibitem{mrg97}
E.H. Morgan, R.A. Remillard, and J. Greiner, 1997, \emph{ApJ}, 482, 993

\bibitem{nes03}
E. Nespoli, et al., 2003, \emph{A\&A}, 412, 235

\bibitem{now05}
M.A. Nowak, et al., 2005, \emph{ApJ}, 626, 1006

\bibitem{rem99}
R.A. Remillard, et al., 1999, \emph{ApJ}, 522, 397

\bibitem{rod}
J. Rodriguez, et al., 2004, \emph{ApJ}, 615, 416

\bibitem{vig03}
F. Vignarca, et al., 2003, \emph{A\&A}, 397, 729

\bibitem{wij}
R. Wijnands, and M van der Klis, 1999, \emph{ApJ},  514, 939

\bibitem{zdz}
A. Zdziarski, et al., 2001, \emph{ApJ}, 554, L48

\end{thebibliography}
\end{document}